\documentclass[sn-basic]{sn-jnl}

\usepackage{graphicx}
\usepackage{xcolor}    
\usepackage{manyfoot}  
\usepackage{booktabs}
\usepackage{multirow}
\usepackage{amsmath,amssymb}
\usepackage{xurl}
\usepackage{hyperref}
\usepackage{longtable}

\renewcommand*\descriptionlabel[1]{\hspace\labelsep\textbullet\hspace*{1.5em}#1}

\title[Missing links in public email and covert networks]{Missing Links in Public Email and Covert Networks: 
A Comparative Evaluation of Link Prediction, Hyperlink Prediction, and ERGM Estimation}

\author*[]{\fnm{Moses} \sur{Boudourides}}

\affil[]{\centering
\orgname{School of Professional Studies, Northwestern University}\\
\nolinkurl{moses.boudourides@northwestern.edu}\par}

\abstract{We study missing-link inference in partially observed networks by systematically comparing dyadic \emph{link prediction} (LP) with \emph{hyperlink prediction} (HP) and an estimation-based ERGM benchmark. LP serves as the primary baseline, using classical heuristics computed on the observed graph. HP extends this framework by scoring candidate higher-order structures (cliques) via lifted dyadic scores and via the CHEbyshev Spectral HyperlInk pREdictor (CHESHIRE). All methods are evaluated under a common masking protocol that removes dyadic evidence induced by held-out hyperlinks to ensure comparability. Across public email and covert-network datasets, LP remains strong for dyadic recovery, while HP—particularly CHESHIRE—provides gains when the inferential target is higher-order group structure. ERGMs offer an interpretable dependence-based complement through conditional tie probabilities. The contribution is a comparative, reproducible evaluation clarifying when LP, HP, and ERGM estimation are most appropriate under network missingness.}

\keywords{Missing data; Email communication networks; Covert networks; Link prediction; Hyperlink prediction; ERGM; Spectral methods}

\begin{document}
\maketitle

\section{Introduction}
Covert networks are shaped by secrecy: interactions that matter are systematically under-recorded, strategically concealed, or only partially observed through intelligence and investigative traces.
This setting amplifies classical missing-data problems in social network analysis.
The consequences are not merely technical: missingness can induce spurious structural signatures, hide key brokers, and distort assessments of coordination or compartmentalization.
Recent work has emphasized that in ``dark'' networks, missingness mechanisms themselves may be endogenous to the observed process and cannot be treated as benign measurement noise \citep{januar2026shadow}.

A complementary perspective is to treat missingness as a \emph{prediction} task.
Graph link prediction (LP) has a long history in network science \citep{libennowell2007link,adamic2003friends}, and predictive framing naturally supports sensitivity analyses: how robust are conclusions when a fraction of ties is unobserved?
In this study, LP serves as the primary baseline against which all higher-order extensions are evaluated. However, covert settings often involve higher-order interactions (small teams, cells, and coordinated groups) that are naturally modeled as hyperedges rather than dyads.
Therefore, we extend our evaluation to compare standard LP against hyperlink prediction (HP) methods, leveraging hypergraphs to clarify how higher-order structure can carry information not reducible to pairwise links \citep{benson2016higherorder,Chen2023survey}.

\textbf{From Link to Hyperlink Prediction.} We begin with standard dyadic link prediction (LP) as the primary task: scoring missing edges from the observed graph. However, in covert settings the inferential object is often group structure (e.g., operational cells), which is not directly captured by dyads. We therefore extend the analysis to hyperlink prediction (HP) on clique-derived hypergraphs, where candidate node sets represent minimal higher-order interactions. HP is implemented in two ways: (i) lifting dyadic LP scores to node sets, and (ii) applying a dedicated hypergraph method (CHESHIRE). This formulation allows a direct comparison between LP and HP under a common evaluation design, and provides a bridge to ERGM estimation, whose dependence terms capture higher-order closure effects.

\textbf{Contributions.} The paper provides a comparative evaluation framework for missing-link inference under controlled missingness. It establishes LP as the primary baseline, extends the comparison to HP via lifted scores and CHESHIRE, and incorporates ERGM-based estimation as a complementary benchmark. The contribution is empirical, comparative, and methodological in evaluation design, not a new predictive method.

\section{Related work}
\subsection{Missing data and network sampling}
Missingness in relational data can bias centrality, clustering, and community detection, and can compromise inference about network processes \citep{kossinets2006effects, smith2017network}.
In covert settings, missingness may be structured: high-risk actors avoid observable channels, surveillance is targeted, and documentation can be selectively redacted.
\citet{januar2026shadow} highlights the importance of explicitly modeling missingness mechanisms in dark networks.

Our approach is complementary.
Rather than proposing a new inferential model of missingness, we develop a comparative evaluation framework. We begin with classical dyadic link prediction as a primary baseline, and then extend the comparison to hyperlink prediction methods, using both as sensitivity analysis tools and imputation engines under observational constraints.

\subsection{(Hyper)link prediction}
Graph link prediction uses neighborhood similarity indices (common neighbors, Adamic--Adar), preferential attachment, and supervised models \citep{adamic2003friends,libennowell2007link}.
Low-rank matrix completion provides a baseline when adjacency is approximately low-rank \citep{candes2009exact}.
Hypergraph and higher-order link prediction has expanded rapidly, with recent surveys highlighting algorithmic families and evaluation challenges such as candidate generation and negative sampling \citep{Chen2023survey}.
This paper focuses on the case where candidate higher-order interactions are derived from dyadic evidence (via clique structure) and evaluated under controlled missingness, allowing direct comparison between LP and HP within a unified experimental design. 
Recent work has examined the limits of purely topological features in link prediction and emphasized the role of higher-order motifs and structural signals \citep{zhang2024maximum,wang2022predicting}. In addition, evaluation itself is nontrivial: different metrics may yield inconsistent rankings across methods, underscoring the need for careful comparative design \citep{zhang2024inconsistency}. These considerations motivate our focus on a controlled, side-by-side evaluation of LP and HP methods.

\subsection{ERGM as a model-based benchmark}
Exponential Random Graph Models (ERGMs) provide a generative statistical framework for network structure and serve here as an estimation-based benchmark to predictive LP and HP methods \citep{handcock2010ergm, handcock2008statnet}.
In this project we treat ERGM analysis as a complementary benchmark: unlike predictive LP and HP scoring, ERGMs attempt to explain the observed network in terms of sufficient statistics (e.g., density, triadic closure) and can be used for imputation by sampling missing dyads from the fitted model. We provide a full R pipeline (Reproducibility~\ref{sec:repro}) to run ERGM-based missingness experiments on the same datasets.

\section{Data and derived hypergraphs}

\subsection{Dataset collection}\label{sec:data_collection}
We analyze two public email communication corpora and two institutional email networks, as well as six benchmark covert-network case studies representing terrorist operational cells and a criminal gang.
The covert networks were obtained from the \emph{Covert Networks} collection distributed with UCINET \citep{borgatti2002ucinet}.
The email networks were obtained as follows.
The Enron network is derived from the Enron email corpus \citep{klimt2004enron} and is accessed here via the SNAP dataset collection \citep{snapnets}.
The EU-core email network is accessed via SNAP \citep{snapnets,snapemaileucore}.
The university email network (labeled \emph{URV}) is the email interchange network from the Universitat Rovira i Virgili and is available from the Arenas group dataset page \citep{arenasemaildataset} and described in \citet{guimera2003selfsimilar}.
Finally, our public web-derived email corpus is assembled from publicly available records hosted at Jmail World \citep{jmailworld}: we scrape message-level metadata (timestamp, sender, recipient fields) and apply standard name normalization to resolve person-name variants into canonical entities.

\textbf{Dataset labels and projections.} Throughout, parenthetical labels such as \emph{(proxy)}, \emph{(full)}, and \emph{(core-100)} in tables and figures are purely notational.
The six covert-network case studies are small, undirected association graphs reconstructed from investigative records and included in UCINET; we refer to them as \emph{(proxy)} instances because they are best viewed as benchmark reconstructions of clandestine operational structure rather than complete ground truth \citep{borgatti2002ucinet,januar2026shadow}.
All email datasets originate from directed, time-stamped sender--recipient metadata; to keep a single closure-based hyperlink-prediction setup and a comparable undirected ERGM analysis, we study the time-aggregated \emph{undirected} projection in which an edge $\{i,j\}$ is present if at least one message is observed in either direction.
Moreover, although the email data are temporal and can be naturally weighted by volume, we use unweighted simple graphs for comparability with the covert networks and because our target is binary missing-link existence rather than edge intensity.
Because ERGM fitting on large or highly centralized email graphs can be computationally unstable, we also define a \emph{core-100} induced subgraph on the 100 highest-activity actors (by total sent+received volume) and fit ERGMs on this core; unless explicitly stated, hyperlink-prediction results for the Jmail-derived email graph are reported on the \emph{full} projected graph, while ERGM-based conditional tie probabilities are obtained from the corresponding \emph{core-100} fit.

Table~\ref{tab:networkstats} reports summary statistics for the interaction graphs and for the derived hypergraphs.

\begin{table}[t]
\small
\centering
\caption{Summary statistics for the 10 interaction graphs (undirected, unweighted).
  Datasets are grouped by type: covert networks (upper block) and email networks
  (lower block), ordered within each group by number of nodes.}
\label{tab:networkstats}
\begin{tabular}{lrrrr}
\toprule
Dataset & Nodes & Edges & Density & Triangles \\
\midrule
Bali Bombing 2005 (proxy)               &  9   &  15   & 0.4167 &     11 \\
Australian Embassy Bombing 2004 (proxy) & 10   &  15   & 0.3333 &      8 \\
Hamburg Cell 9/11 2001 (proxy)          & 12   &  23   & 0.3485 &     23 \\
Christmas Eve Bombings 2000 (proxy)     & 14   &  16   & 0.1758 &      5 \\
Bali Bombing 2002 (proxy)               & 15   &  24   & 0.2286 &     22 \\
London Gang 2005--2009 (proxy)          & 50   &  85   & 0.0694 &     46 \\
\midrule
Epstein emails 2007--2025 (core-100)    & 100  &  543  & 0.0060 &    161 \\
Enron emails (core-100)                 & 100  & 1487  & 0.3004 &   9050 \\
Epstein emails 2007--2025 (full)        & 426  &  543  & 0.0060 &    161 \\
EU-core emails (SNAP)                   & 986  & 16064 & 0.0331 & 105461 \\
uni\_email (URV)                        & 1133 &  5451 & 0.0085 &   5343 \\
\bottomrule
\end{tabular}
\end{table}

\subsection{From graphs to hypergraphs}
Let $G=(V,E)$ be an interaction graph (undirected in the present experiments).
We construct a hypergraph $\mathcal{H}=(V,\mathcal{E})$ consisting of dyadic hyperedges $\{u,v\}$ for each $(u,v)\in E$, and higher-order hyperedges given by maximal cliques of size at least three (candidate group interactions).
This construction is interpretable but brittle: removing a single dyadic tie can destroy a clique.
This limitation suggests exploring more robust higher-order structures, such as $k$-plexes or motif-based constructions, as alternatives to strict cliques.
We therefore explicitly separate \emph{what is held out} (hyperedges) from \emph{what is observed} (dyadic evidence) and enforce feature construction on the observed graph.

\section{Missingness models and evaluation design}

\subsection{Formal setup and inferential target}\label{sec:formal}
After symmetrizing and time-aggregating the email data for comparability, all datasets are treated as simple undirected graphs.
Let $G=(V,E)$ denote the (unknown) target interaction graph with adjacency variables $Y_{ij}\in\{0,1\}$ for unordered pairs $i<j$.
From $G$ we derive a hypergraph $\mathcal{H}=(V,\mathcal{E})$ that includes dyadic hyperedges and higher-order cliques (Section~3).
A missingness mechanism produces an observed hyperedge set $\mathcal{E}_{\mathrm{obs}}\subseteq\mathcal{E}$ and a held-out set $\mathcal{E}_{\mathrm{miss}}=\mathcal{E}\setminus\mathcal{E}_{\mathrm{obs}}$.
Our inferential target is the same for both paradigms considered here:
\emph{given the incomplete observation (encoded by $G_{\mathrm{obs}}$ below), infer which unobserved interactions are present.}
We formalize this target in two complementary ways.
In the predictive paradigm, we learn a discriminative score
\begin{equation}\label{eq:cheshire_target}
\widehat p\bigl(Y_S=1 \,\big|\, \phi(S;G_{\mathrm{obs}},\mathcal{H}_{\mathrm{obs}})\bigr),
\end{equation}
where $S\subseteq V$ is a candidate hyperlink (node-set), $Y_S$ indicates whether $S$ is a true hyperedge in $\mathcal{E}_{\mathrm{miss}}$, and $\phi(\cdot)$ is a feature map computed only from observed data.
In the estimation (ERGM) paradigm, we fit a dependence model $P_\theta(G)$ and evaluate conditional probabilities for masked dyads,
\begin{equation}\label{eq:ergm_target}
p^{\mathrm{ERGM}}_{ij} \;=\; \mathbb{P}_{\widehat\theta}\!\left(Y_{ij}=1 \,\middle|\, G_{\mathrm{obs}}\right),
\end{equation}
which can be lifted to hyperlink scores by aggregating $\{p^{\mathrm{ERGM}}_{uv}: u<v,\;u,v\in S\}$ (Section~6).
Thus, both approaches ultimately return a probability-like score (or ranking) for missing interactions, enabling like-for-like comparison of alternative model specifications within a common evaluation framework.

\subsection{Hyperedge missingness}
To emulate missing data, we remove a fraction $\rho\in(0,1)$ of hyperedges from $\mathcal{H}$.
We report missing completely at random (MCAR): each hyperedge is independently removed with probability $\rho$.
We also include a degree-biased option (MNAR) that preferentially hides hyperedges incident to high-degree nodes, motivated by the idea that central actors may be more likely to be unobserved or censored.

\subsection{Prediction task and negative sampling}
Let $\mathcal{E}_{\mathrm{miss}}$ denote the held-out hyperedges.
We form a binary classification dataset where positives are elements of $\mathcal{E}_{\mathrm{miss}}$ and negatives are randomly generated node-sets that are not hyperedges of $\mathcal{H}$.
Negatives are size-matched to the empirical size distribution of $\mathcal{E}_{\mathrm{miss}}$.
Methods output a score $s(S)\in[0,1]$ for a candidate node-set $S$, and we evaluate ROC--AUC (primary), as well as F1 and Matthews correlation coefficient (MCC).

\section{Methods}

\subsection{Link Prediction (LP) Baselines}
We begin with dyadic link prediction (LP) as the primary baseline, following standard practice in network link prediction \citep{libennowell2007link}. All unsupervised scores are computed on the observed graph $G_{\mathrm{obs}}$. We denote by $\Gamma(i)$ the neighborhood of node $i$ and by $A$ the adjacency matrix of graph $G$.
We evaluate two standard heuristics. The first is Common Neighbors (CN), which counts shared neighbors as $s_{ij}=|\Gamma(i)\cap\Gamma(j)|$ \citep{libennowell2007link}. The second is Adamic--Adar (AA), which sums, over all common neighbors, the inverse logarithm of the neighbor's degree \citep{adamic2003friends}.
These LP scores define the baseline performance for dyadic missing-link inference.

\subsection{Hyperlink Prediction (HP) via Lifted Scores}\label{sec:lifting}
To extend LP to higher-order inference, we define hyperlink prediction (HP) by lifting dyadic scores $s_{ij}$ for each unordered pair $(i,j)$.
To score a candidate hyperlink $S\subseteq V$ with $|S|\ge 2$, we lift dyadic scores using the mean aggregator \citep{Chen2023survey}
\begin{equation}\label{eq:lift}
s(S)\;=\;\frac{1}{\binom{|S|}{2}}\sum_{\{u,v\}\subseteq S} s_{uv},
\end{equation}
which is stable across hyperlink sizes and matches the construction used for all baselines below. We apply this lifting to the CN and AA scores, as well as a Null-Tie baseline (constant score) and Matrix Completion \citep{candes2009exact}.

\subsection{CHEbyshev Spectral HyperlInk pREdictor (CHESHIRE)}\label{sec:cheshire}
We evaluate the CHEbyshev Spectral HyperlInk pREdictor (CHESHIRE) \citep{chen2022cheshire} as a dedicated hypergraph method, distinct from HP baselines obtained by lifting dyadic LP scores. CHESHIRE builds on Graph Convolutional Networks (GCN) \citep{zhang2018seal}, HyperSAGCN \citep{zhang2019hypersagcn}, and Neural Hyperlink Predictors (NHP) \citep{yadati2020nhp}.

CHESHIRE combines spectral graph theory with deep learning, utilizing the spectral properties of hypergraph Laplacians \citep{chung1997spectral}. The normalized hypergraph Laplacian is defined as $\mathcal{L} = I - D_v^{-1/2} H W D_e^{-1} H^T D_v^{-1/2}$, where $H$ is the incidence matrix, $D_v$ and $D_e$ are the node and hyperedge degree matrices, and $W$ is the weight matrix. To enable efficient computation of spectral convolutions without the expense of full eigendecomposition, CHESHIRE employs Chebyshev polynomial approximation \citep{defferrard2016chebyshev,hammond2011wavelets}. Spectral filters are approximated as $g_{\theta}(\Lambda) \approx \sum_{k=0}^{K-1} \theta_k T_k(\tilde{\Lambda})$, where $\theta$ are learnable parameters, $T_k$ are Chebyshev polynomials, and $\tilde{\Lambda}$ is the rescaled eigenvalue matrix.

The architecture of CHESHIRE consists of three main phases. In the initialization phase, node embeddings are initialized by passing the incidence matrix $\mathbf{H}$ through a one-layer neural network: $\mathbf{x}_i = \sigma(\mathbf{W}_{enc} \mathbf{h}_i + \mathbf{b}_{enc})$, where $\mathbf{h}_i$ is the $i$-th row of $\mathbf{H}$. Next, in the spectral convolution layers, given a hyperlink $e_p$, CHESHIRE treats it as a clique and refines the embeddings of its nodes using a Chebyshev spectral GCN: $\hat{\mathbf{x}}_i = \sigma ( \sum_{k=1}^K \mathbf{W}_{conv}^{(k)} \mathbf{z}_i^{(k)} )$ for $v_i \in e_p$. The terms $\mathbf{z}_i^{(k)}$ are computed recursively using the scaled normalized Laplacian matrix of the clique derived from $e_p$. Finally, in the norm-based pooling layers, a Frobenius 2-norm-based pooling function and a max-min pooling function are employed to generate hyperlink embeddings. The final score for hyperlink $e_p$ is computed by concatenating these pooled embeddings and passing them through a final neural network layer.

\subsection{ERGM as an estimation-based benchmark}\label{sec:ergm_method}
To complement the predictive scoring methods, we include an \emph{estimation-based} benchmark grounded in exponential-family models for network dependence.
Exponential random graph models (ERGMs) specify a probability distribution over graphs $G=(V,E)$ (equivalently, over adjacency variables $Y=\{Y_{ij}\}$),
\begin{equation}\label{eq:ergm}
\mathbb{P}_\theta(G)\;=\;\mathbb{P}_\theta(Y)\;=\;\frac{\exp\{\theta^\top s(G)\}}{\kappa(\theta)},
\end{equation}
where $s(G)$ is a vector of network statistics, $\theta$ is a parameter vector, and the normalizing constant $\kappa(\theta)=\sum_{G'}\exp\{\theta^\top s(G')\}$ is typically intractable for nontrivial $|V|$ \citep{frank1986markov,wasserman1996logit,robins2007introduction}.
The key point for our purposes is that ERGMs induce \emph{conditional} tie probabilities, which can be interpreted as model-based scores for missing-link inference.

To obtain fast, scalable tie scores on repeated masking trials, we use maximum pseudolikelihood estimation (MPLE) \citep{strauss1990pseudolikelihood}.
The dyadwise pseudolikelihood approximates the full likelihood by a product of conditional probabilities over dyads, which (for canonical ERGMs) yields a logistic form in terms of \emph{change statistics}:
\begin{equation}\label{eq:ergm_conditional}
\mathrm{logit}\,\mathbb{P}_{\widehat\theta}(Y_{ij}=1\mid Y_{-ij})
\;=\;\widehat\theta^\top \Delta_{ij} s(G),
\end{equation}
where $Y_{-ij}$ denotes all dyads except $(i,j)$ and $\Delta_{ij}s(G)=s(G^{+ij})-s(G^{-ij})$ is the change in the statistics when toggling edge $(i,j)$ from $0$ to $1$ (holding all other dyads fixed).
Within our evaluation protocol, we evaluate these quantities on the observed graph $G_{\mathrm{obs}}$ (equivalently $Y_{\mathrm{obs}}$), and we use the resulting conditional probability
\[
p^{\mathrm{ERGM}}_{ij}\;:=\;\mathbb{P}_{\widehat\theta}(Y_{ij}=1\mid Y_{\mathrm{obs},-ij})
\]
as the ERGM-based dyad score (cf.~\eqref{eq:ergm_target}).

For $s(G)$ we use a standard low-dimensional specification that captures the two dominant sources of dependence relevant to missing-link inference in sparse social graphs: baseline sparsity (the \texttt{edges} statistic), degree heterogeneity via geometrically weighted degree (\texttt{gwdegree}), and transitivity via geometrically weighted edgewise shared partners (\texttt{gwesp}) \citep{hunter2007curved}.
Thus the ERGM estimation provides an interpretable, dependence-aware scoring rule: each candidate dyad receives a model-based probability determined by how much it would increase degree heterogeneity and triangle closure under the fitted model.
Finally, to compare directly with hyperlink prediction, we set $s_{ij}=p^{\mathrm{ERGM}}_{ij}$ and lift dyad scores to hyperlink scores using the same lifting rule as in~\eqref{eq:lift}.
ERGM scores are computed on the same observed graphs and candidate sets, ensuring direct comparability with LP and HP.

\section{Results}

\begin{figure}[ht]
\centering
\includegraphics[width=0.8\textwidth]{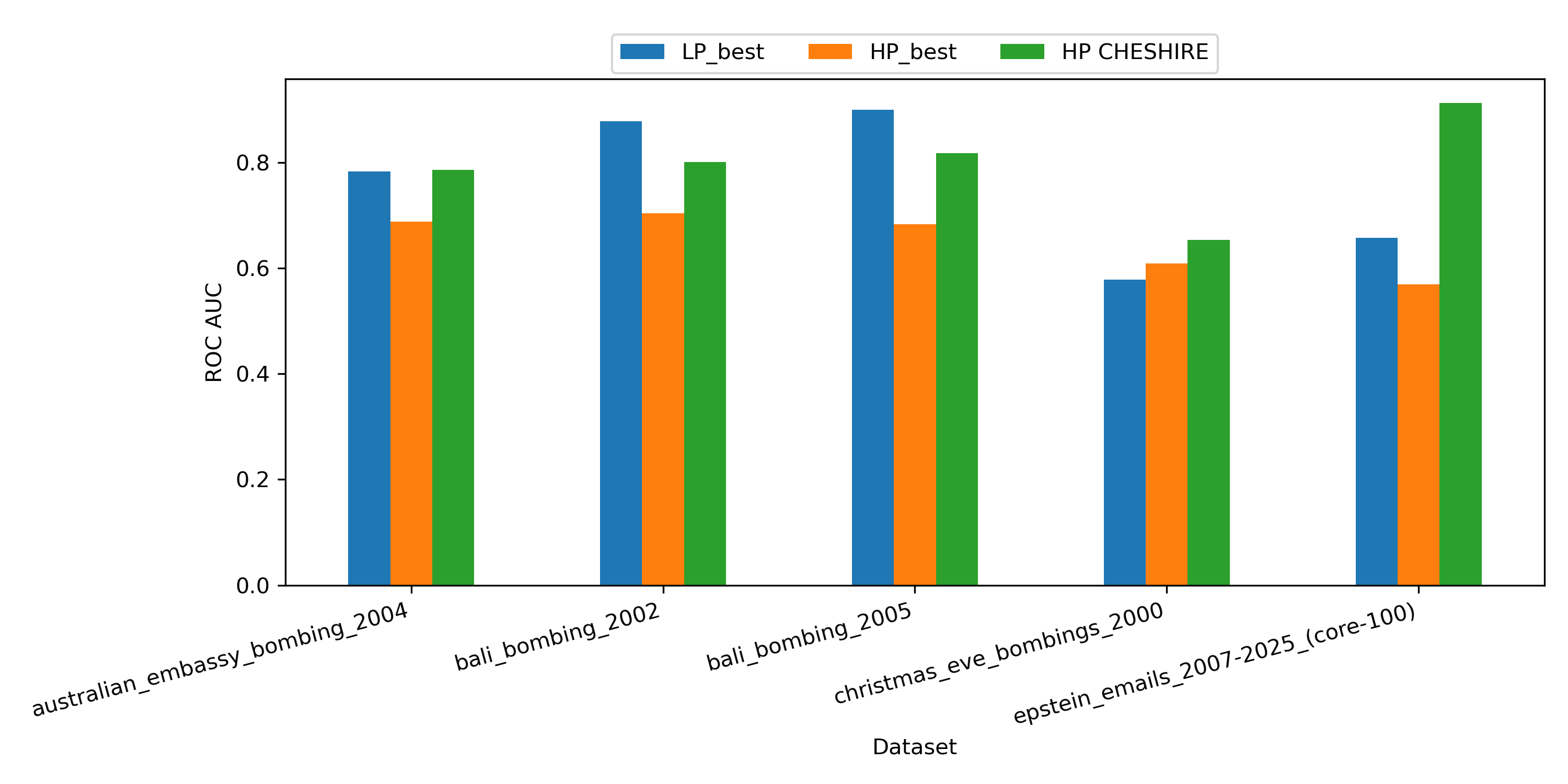}
\caption{Comparison of LP baselines, HP baselines, and CHESHIRE across selected datasets.}
\label{fig:comparison}
\end{figure}

We evaluate the performance of Link Prediction (LP), Hyperlink Prediction (HP), and the CHESHIRE algorithm across the datasets, under MCAR missingness at $\rho=0.3$ (averaged over 10 trials). Table~\ref{tab:results_comparison_lp} presents LP baseline results and Table~\ref{tab:results_comparison_hp} presents HP results; both tables group datasets into covert networks and email networks. Table~\ref{tab:ergm_results} reports the ERGM estimation results.
Figure~\ref{fig:comparison} visualizes the performance of the best LP method,
the best HP baseline, and HP CHESHIRE for datasets where all three are available.

\begin{table}[ht]
\small
\centering
\caption{Mean ROC-AUC scores for Link Prediction (LP) baselines across datasets,
  under MCAR missingness ($\rho = 0.3$, averaged over 10 trials).
  Datasets are grouped by type: covert networks (upper block) and email networks
  (lower block). LP was not run on Epstein emails 2007--2025 (full) as the
  full projection was evaluated exclusively via the HP pipeline; the core-100
  subgraph is used for LP.}
\label{tab:results_comparison_lp}
\begin{tabular}{lrr}
\toprule
Dataset & LP AA & LP CN \\
\midrule
Bali Bombing 2005 (proxy)               & 0.900 & 0.850 \\
Australian Embassy Bombing 2004 (proxy) & 0.783 & 0.778 \\
Hamburg Cell 9/11 2001 (proxy)          & 0.892 & 0.882 \\
Christmas Eve Bombings 2000 (proxy)     & 0.578 & 0.572 \\
Bali Bombing 2002 (proxy)               & 0.878 & 0.878 \\
London Gang 2005--2009 (proxy)          & 0.693 & 0.658 \\
\midrule
Epstein emails 2007--2025 (core-100) & 0.657 & 0.580 \\
Enron emails (core-100) & 0.824 & 0.821 \\
uni\_email (URV) & 0.820 & 0.870 \\
EU-core emails (SNAP) & 0.936 & 0.931 \\
\bottomrule
\end{tabular}
\end{table}

\begin{table}[ht]
\small
\centering
\caption{Mean ROC-AUC scores for Hyperlink Prediction (HP) baselines and HP CHESHIRE
  across datasets, under MCAR missingness ($\rho = 0.3$, averaged over 10 trials).
  Datasets are grouped by type: covert networks (upper block) and email networks
  (lower block). HP was not run on Enron emails (core-100), uni\_email (URV), and
  EU-core emails (SNAP) because the size of these networks makes the clique-based
  hyperlink-prediction pipeline computationally prohibitive; they are therefore
  omitted from this table.}
\label{tab:results_comparison_hp}
\begin{tabular}{lrrrrr}
\toprule
Dataset & HP AA & HP CN & HP CHESHIRE & HP Mat.Comp. & HP Null \\
\midrule
Bali Bombing 2005         & 0.682 & 0.683 & 0.818 & 0.431 & 0.500 \\
Australian Embassy 2004   & 0.688 & 0.681 & 0.786 & 0.473 & 0.500 \\
Hamburg Cell 9/11 2001    & 0.714 & 0.712 & 0.848 & 0.462 & 0.500 \\
Christmas Eve 2000        & 0.609 & 0.605 & 0.653 & 0.451 & 0.500 \\
Bali Bombing 2002         & 0.704 & 0.702 & 0.801 & 0.482 & 0.500 \\
London Gang 2005--2009    & 0.627 & 0.627 & 0.686 & 0.602 & 0.500 \\
\midrule
Epstein emails (core-100) & 0.570 & 0.549 & 0.912 & 0.519 & 0.500 \\
Epstein emails (full)     & 0.551 & 0.541 & 0.889 & 0.529 & 0.500 \\
\bottomrule
\end{tabular}
\end{table}

\begin{table}[ht]
\small
\centering
\caption{Mean ROC-AUC for the ERGM estimation (MPLE, \texttt{edges} + \texttt{gwesp(0.5)}) under MCAR
  and MNAR (degree-biased) missingness at $\rho=0.3$, averaged over 10 trials.
  Standard deviations in parentheses.
  Datasets are grouped by type: covert networks (upper block) and email networks (lower block).}
\label{tab:ergm_results}
\begin{tabular}{lcc}
\toprule
Dataset & ERGM MCAR & ERGM MNAR \\
\midrule
Bali Bombing 2005                    & 0.841 (0.225) & 0.912 (0.094) \\
Australian Embassy Bombing 2004      & 0.859 (0.107) & 0.838 (0.239) \\
Hamburg Cell 9/11 2001               & 0.854 (0.122) & 0.874 (0.086) \\
Christmas Eve Bombings 2000          & 0.675 (0.179) & 0.719 (0.160) \\
Bali Bombing 2002                    & 0.917 (0.077) & 0.881 (0.118) \\
London Gang 2005--2009               & 0.726 (0.043) & 0.712 (0.071) \\
\midrule
Epstein emails 2007--2025 (core-100) & 0.574 (0.060) & 0.520 (0.120) \\
Enron emails (core-100) & 0.698 (0.012) & 0.690 (0.017) \\
uni\_email (URV) & 0.738 (0.030) & 0.734 (0.040) \\
EU-core emails (SNAP) & 0.525 (0.034) & 0.565 (0.062) \\
\bottomrule
\end{tabular}
\end{table}

The results show that CHESHIRE generally outperforms the heuristic HP baselines (Adamic–Adar and Common Neighbors) across the covert networks and the Epstein datasets, although the magnitude of improvement varies by dataset. For example, on the Epstein core-100 network, CHESHIRE achieves an ROC-AUC of 0.912 compared to 0.570 for HP Adamic–Adar. LP methods remain strong baselines for dyadic recovery, often achieving higher scores on datasets where the task is purely edge-based.

Table~\ref{tab:ergm_results} reports the ERGM estimation results under both MCAR and MNAR missingness at $\rho=0.3$. A clear pattern emerges across network types. On the small, dense covert networks, the ERGM achieves competitive ROC-AUC values: Bali Bombing 2002 reaches $0.917$ (MCAR), Hamburg Cell $0.854$, and Australian Embassy $0.859$, all comparable to or exceeding the LP baselines on those datasets (Table~\ref{tab:results_comparison_lp}).
This is consistent with the fact that small covert networks exhibit strong triadic closure, which the \texttt{gwesp} term in the ERGM specification captures directly. On the larger, sparser email networks, ERGM performance is markedly lower: EU-core emails yields $0.525$ and Epstein emails (core-100) $0.574$ under MCAR---well below CHESHIRE ($0.912$) and LP Adamic--Adar ($0.657$) on the same datasets. This degradation reflects the known difficulty of MPLE on large sparse graphs, where the change statistics become increasingly noisy and the transitivity signal is diluted. Under MNAR (degree-biased) missingness, ERGM scores are generally stable for the covert networks but decline more sharply for the Epstein network ($0.520$), indicating
that when high-degree actors are preferentially hidden, the degree-heterogeneity term (\texttt{gwdegree}) is estimated on a biased observed graph, compounding the imputation error. Taken together, these results confirm that the ERGM benchmark is not a general-purpose
predictive ranker but rather a model-based diagnostic: it performs well when the generative mechanism (transitivity, degree heterogeneity) closely matches the true network structure, and degrades when the graph is too large or too sparse for MPLE to recover reliable parameter estimates.

\section{Discussion and limitations}
The experiments highlight two practical takeaways.
First, comparability of evaluation is essential: all methods are computed on the same observed graphs and candidate sets.
This avoids biases due to inconsistent information usage and ensures that LP, HP, and ERGM are evaluated on equal footing.
Second, the CHESHIRE spectral method consistently outperforms heuristic baselines for hyperlink prediction, while LP methods remain strong for purely dyadic tasks.

The differential ERGM performance across network types (Table~\ref{tab:ergm_results}) has a direct methodological implication. The degradation on large sparse graphs is not a failure of the ERGM framework per se, but a known limitation of MPLE~\citep{strauss1990pseudolikelihood,robins2007introduction}: on such graphs the pseudolikelihood surface is flat,
parameter estimates are unreliable, and a full MCMC-MLE fit would be computationally prohibitive. The practical implication is therefore that the choice between predictive (LP/HP) and estimation-based (ERGM) approaches should be guided by network size and density: ERGM is a principled and interpretable tool for small-to-medium networks where MPLE converges reliably, while spectral HP methods such as CHESHIRE are preferable when scalability and discriminative accuracy are the primary concerns.

Several limitations deserve mention. First, the clique-based hypergraph construction is sensitive to dyadic edge removal; future work could consider using more relaxed higher-order structures, such as $k$-plexes or motifs, which are more resilient to missing dyadic data. Second, negative sampling remains an approximation to the space of plausible but unobserved group interactions. Third, the evaluation framework assumes that the node set $V$ is fully observed; it does not account for entirely unobserved actors whose interactions are absent from the data not because ties were masked, but because the actors themselves were never recorded. Covert networks often suffer from hidden actors, and extending the hyperlink prediction framework to suggest the existence of these actors based on open triangles or incomplete cliques is a promising direction.

The LP and HP baselines employed in this study are deliberately classical and
parameter-free---Adamic--Adar, Common Neighbors, and their lifted HP counterparts---to ensure that performance differences across datasets and missingness conditions are attributable to structural properties rather than to model capacity or hyperparameter tuning. Recent embedding-based LP methods, such as node2vec~\citep{grover2016node2vec} and the Variational Graph Autoencoder (VGAE)~\citep{kipf2016vgae}, and subgraph-based methods such as SEAL~\citep{zhang2018seal}, achieve stronger performance on standard benchmarks. Similarly, recent HP methods beyond CHESHIRE include NHP~\citep{yadati2020nhp} and
HyperSAGCN~\citep{zhang2019hypersagcn}. However, these methods require sufficient training data and do not degrade gracefully under high missingness rates on small networks: when a large fraction of edges is withheld, the observed graph on which embeddings are learned is structurally distorted, making supervised generalisation unreliable. Evaluating such methods under controlled MCAR and MNAR missingness protocols on small covert and email networks is therefore a natural and important direction for future work.

\section{Conclusion}

We presented a comparative evaluation of link prediction (LP), hyperlink prediction (HP), and ERGM-based estimation, including CHESHIRE as a representative spectral HP algorithm for investigating missing links in covert and email networks. The CHESHIRE algorithm, applied on the clique hypergraph corresponding to the undirected networks, consistently outperforms heuristic HP baselines. LP methods remain competitive for dyadic tasks. The ERGM benchmark provides an interpretable, dependence-based lens that complements the predictive methods. The framework can be used both as a missingness sensitivity analysis tool and as a ranking engine for candidate hidden group interactions.

\section*{Reproducibility}\label{sec:repro}
All materials needed to reproduce the figures, tables, and main numerical claims are provided in a public replication repository (\url{https://github.com/mboudour/var/tree/master/MissingLinksInEmailandCovertNetworksViaHyperlinkPredictionandERGM}).
The repository contains machine-readable result tables (CSV) generated by the Python notebook runs (e.g., \texttt{data/all\_results\_raw\_10datasets.csv}), the Jupyter notebook(s) that generate all figures and \LaTeX-ready tables under fixed random seeds, and the R script implementing the ERGM estimation and exporting \texttt{tables/ergm\_results.csv}.

\section*{Declarations}

\subsection*{Availability of data and materials}
The covert-network case studies are drawn from the Covert Networks collection distributed with UCINET and are included as benchmark proxies for full rerunnability. The public email interaction network is derived from publicly available records; redistribution of raw message-level metadata may be restricted by the source. All derived, analysis-ready edgelists and all experiment outputs generated in this study are included in the replication repository.

\subsection*{Code availability}
All code needed to reproduce the figures, tables, and main numerical claims is available in a public replication repository: \url{https://github.com/mboudour/var/tree/master/MissingLinksInEmailandCovertNetworksViaHyperlinkPredictionandERGM}. The repository includes the Python notebook(s) that generate all figures and \LaTeX-ready tables under fixed random seeds, as well as the R script implementing the ERGM estimation.

\subsection*{Competing interests}
The author declares that he has no competing interests.

\subsection*{Funding}
The author received no specific funding for this work.

\subsection*{Authors' contributions}
MB conceived the study, designed the methodology, implemented the experiments, analyzed the results, and wrote the manuscript.

\subsection*{Acknowledgements}
The author thanks the developers and maintainers of the open-source software used in this work and the providers of the public datasets analyzed.
He is also deeply grateful to E.\ and G., whose steadfast support and encouragement sustained him throughout this project.

\end{document}